\documentclass[showpacs]{revtex4}
\usepackage{epsfig}
\usepackage{amsmath}
\usepackage{amsfonts}

\newcommand{\qF}{\mathcal{F}}
\newcommand{\qT}{\mathcal{T}}
\newcommand{\qe}{\varepsilon}

\newcommand{\qo}{\omega}
\def\rvec{{\mathbf{r}}}
\def\wvec{{\mathbf{w}}}
\def\dvec{{\mathbf{d}}}
\def\evec{{\mathbf{e}}}
\def\cvec{{\mathbf{c}}}
\def\hvec{{\mathbf{h}}}
\def\Fvec{{\mathbf{F}}}
\def\Tvec{{\mathbf{T}}}
\def\Nvec{{\mathbf{N}}}
\def\Rvec{{\mathbf{R}}}

\begin{document}

\title{Coupling between static friction force and torque for a tripod}

\author{S\'{\i}lvio R. Dahmen}
\affiliation{Instituto de F\'{\i}sica da UFRGS, CP 15051,
90501--970 Porto Alegre RS, Brazil}
\author{Haye Hinrichsen}
\affiliation{Fakult\"at f\"ur Physik und Astronomie, Universit\"at
W\"urzburg, D-97074 W\"urzburg, Germany}
\author{Andrey Lysov}
\affiliation{Department of Physics, Universit\"at
Duisburg-Essen, D-47048 Duisburg, Germany}
\author{Dietrich E. Wolf}
\affiliation{Department of Physics, Universit\"at
Duisburg-Essen, D-47048 Duisburg, Germany}

\pacs{46.55.+d, 81.40.Pq, 45.70.-n, 81.05.Rm}
\keywords{friction; tribology}

\parskip 2mm

\begin{abstract}
If a body is resting on a flat surface, the maximal static friction force before motion sets
in  is reduced if an external torque is also applied. The coupling
between the static friction force and static friction torque is nontrivial as
our studies for a tripod lying on horizontal flat surface show. In this article we report on a series of experiments we performed on a tripod
and compare these with analytical and numerical solutions. It turns out
that the coupling
between force and torque reveals information about the microscopic
properties at the onset to sliding.
\end{abstract}
\pacs{46.55.+d, 81.40.Pq, 45.70.-n, 81.05.Rm}

\maketitle
\section{Introduction}
%
Even though most of the everyday facts about dry macroscopic friction have been
known since the works of Leonardo da Vinci, Amontons, Coulomb and
Euler~\cite{Dowson79}, the physics of static friction, especially
at the onset to sliding,
is not yet fully understood. This applies in particular to bodies
undergoing a simultaneous translational and rotational
motion which, on account of friction, exhibit nontrivial dynamics \cite{ruina,Zeno2004}. In fact, the
sliding friction of a circular disk is reduced if the contact is also spinning
with relative angular velocity $\qo_{\rm n}$ -- a phenomenon which plays
an important role in various games such as curling or ice hockey \cite{voyenli,sh2,nature03}.
It turns out that this reduction depends on the dimensionless ratio
\begin{equation}
\qe = \frac{v_{\rm t}}{\qo_{\rm n} R},
\label{eq:epsilon}
\end{equation}
where $R$ denotes the radius of the disk and $v_{\rm t}$ is the
tangential relative velocity at the center of the contact area.
Based on the Coulomb friction law one obtains a sliding friction force
\begin{equation}
|\mathbf{F}|=\mu_{\rm d} N {\qF}(\qe)
\end{equation}
and a friction torque
\begin{equation}
|\mathbf{T}|=\mu_{\rm d} N R {\qT}(\qe),
\end{equation}
where $\mu_{\rm d}$ is the dynamic friction coefficient and $N$ is
the integrated normal force acting on the contact area. Apart from the limit
of pure sliding $\qe \to \infty$, where ${\qF} \rightarrow 1$,
the functions ${\qF(\qe)}$ and ${\qT(\qe)}$ depend on the pressure
distribution across the contact area. Assuming uniform pressure
over the area of the disk these functions have been evaluated
analytically and compared with experiments \cite{Zeno2004}. An interesting phenomenon occurs in
the case of sliding cylinders: due to an asymmetry in the pressure
exerted on the track, which arises from the torque which
tends to tip the cylinder, an analogon to the Magnus effect appears,
causing the cylinder to describe curved trajectories \cite{Zeno2004,sh3}.

\begin{figure}
\centerline{\epsfig{figure=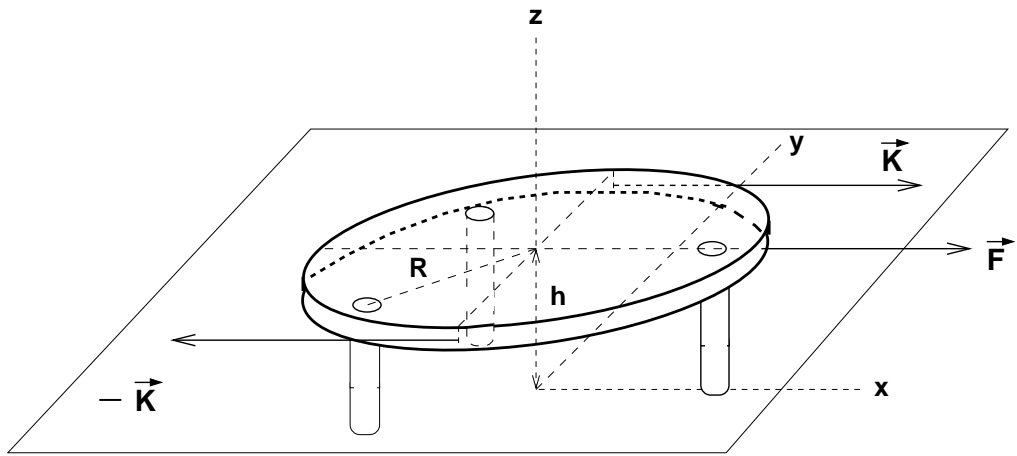,width=0.75\linewidth}}
\caption{The tripod experiment: bolts are evenly arranged around the
perimeter and firmly attached to the disk surface. In the absence of
external forces the load is equally
distributed among the bolts.}
\label{fig:tripod}
\end{figure}

Another class of systems which has been studied extensively in a series of papers
by Shegelski and coworkers is that of sliding tripods with
symmetrically placed legs ($2\pi/3$ radians apart from each other, see Fig. \ref{fig:tripod})
\cite{sh1}.  One observes,
for given initial conditions, trajectories which are serpentine--like (albeit with
low lateral deflection), to curved trajectories with very large lateral deflections (curling).
The translational kinetic energy
may increase and decrease during a full rotation (the overall kinetic energy decreases with time). In this situation one still has a parameter analogous to $\qe$ as in Eq. (\ref{eq:epsilon}) but now there is a nontrivial dependence on $\theta$
which is the angle the leg makes with the horizontal x--axis of some fixed laboratory coordinate system.

The dynamics of rotating and sliding bodies under the action of friction forces is by now fairly well understood.
Not so much
can be said about the {\it statics} of bodies subject to a torque and a force. Here one
important question is the minimal force and torque necessary to set the body
into motion and how these are coupled. That there must be some coupling between them one knows from daily experience:
if a heavy object is to be moved across the floor, it is easier to do so if one applies a torque while pushing it. Some previous analytical studies on disks showed that this is
indeed the case but, contrary to the dynamical case, the static situation depends strongly on the model
one uses for microscopic displacement: due to the inhomogeneity of the local displacement at
each microcontact, some of them are subject to greater lateral stresses than others.
One may then picture two scenarios. In the first one,
a threshold is reached first at those microcontacts where the local stress is maximal.
These then break \textit{irreversibly} and the released stress is
distributed among the remaining microcontacts. As some of these cannot
sustain this increase in stress they also break, triggering an avalanche--like process where
eventually all contacts detach and the disk starts to move. In the other scenario
the broken microcontacts may
immediately rearrange and form new contacts, redistributing the
released stress over the remaining {\it and} the newly
formed contact points. This microscopic stick--slip creeping continues until
all contact points self-organize in such a way that they
sustain on average the same stress. Therefore, by increasing
the external force or torque, all microcontacts of a perfectly rigid slider
reach the threshold of detachment simultaneously.

Recent experimental studies by Rubinstein and coworkers who used photoarrays to monitor the dynamics of contact points at the onset of sliding seem to favor the first scenario \cite{Rubinstein04}. In their experiments a pexiglass slab was pushed in a straigh line over
a macroscopically flat surface. Initially they observed that the
microcontacts give way sequentially, propagating from the trailing edge -- where the force was applied -- to the front edge. The front
starts propagating with about half the Rayleigh speed, accelerates
and then splits up into  a subsonic and an intersonic front. The time scale the contacts give way is extremely short. However, linear
sliding has an intrinsic asymmetry since for an elastic body one expects the microcontacts to break where the force is applied (as the above mentioned experiments confirm) and it is not clear whether this asymmetry is in the last instance responsible for the observed phenomenon.

In this paper we study the interplay of static forces and static torques for a tripod both theoretically and experimentally.
In the next section we present experimental estimates on the
theshold force  and torque needed to set the tripod into motion.
This is followed by a theoretical section, where starting from the idea that all three feet reach the onset of sliding simultaneously, the threshold force and torque are calculated analytically.
We conclude our paper with a discussion of our results and
some perspectives. An
appendix on an exactly solvable case is included at the end.
%
%
\section{Experiments}
\label{ExperimentalSection}
%
In order to determine the critical line of the onset of sliding
we performed a series of experiments where a pulling force and a
torque were applied simultaneously to a tripod standing on a horizontal
surface.

Experiments were carried out using a circular disk made of steel
with radius of $86$ mm. The mass was $1,123$ g. The three feet, which were
mechanically polished, were evenly spaced around the perimeter and placed at a distance of $R=80$ mm from the center
of the disk, to which they were firmly attached. The tripod
was provided with hooks, which were placed at a height of $h=19$ mm above the track
(see Fig.~\ref{fig:tripod}). To measure torques and forces the tripod was placed on a
fixed and macroscopically flat horizontal surface.
Force meters were then 
attached to the hooks.

Once the tripod was on the surface a torque $\mathbf{T}$ (as
indicated in Fig.~\ref{fig:tripod} by the force pair $\mathbf{K}$ and
$-\mathbf{K}$) was applied. The disk was slowly pulled with
force $\mathbf {F}$ until it started moving.
The force meters were set to register the maximum applied pulling
force $\mathbf{F_c}$. For each fixed value of the torque a set of readings for the
maximal force was made. The experiments were repeated several times
under similar temperature and humidity conditions. We note that for
very small pulling forces (large applied torques), the procedure was
inverted: we fixed the force and varied the torque. This is necessary
since around the region where slide sets the range of the torque values becomes extremely
narrow and as it turned out, it was extremely difficult to fix the torque without displacing the disk.
Steel and plastic surfaces were used
and results turned out to be material--independent.

\begin{figure}
\centerline{\epsfig{figure=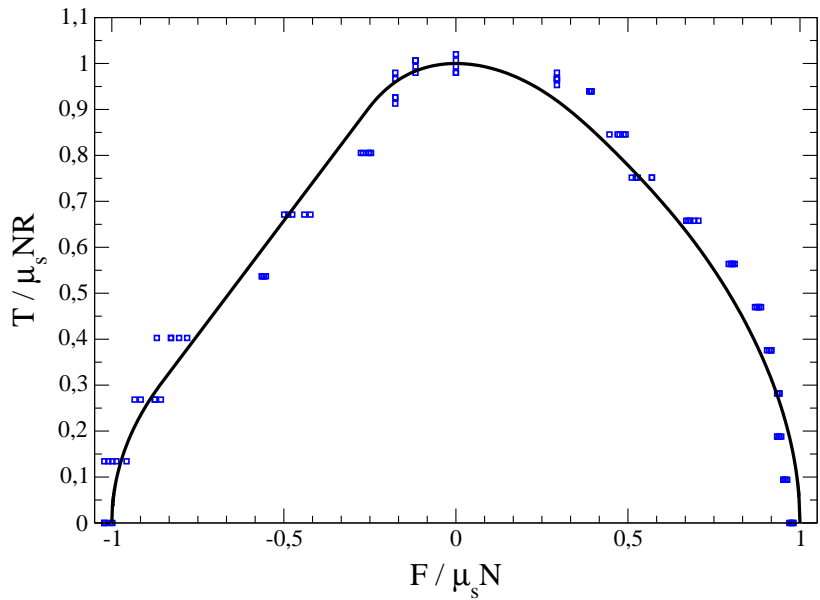,width=0.6\linewidth}}
\caption{Measured values of torque and force for a tripod at the onset of sliding, compared
with the theoretical predictions (solid line; see text for details).
Positive values of $F$ correspond to the direction of the force
as shown in Fig. \ref{fig:tripod}, negative ones to the opposite direction. $\mu_s$ is the static friction coefficient and $N$ the weight of the tripod.}
\label{fig:experiment}
\end{figure}

Contrary to the case of a
disk surface in direct contact with the track, in the tripod configuration the
observed Force $\times$ Torque relationship depends not only on the microscopic
scenario but also on the position of the hook points
relative to the position of the bolts, i.e. on the point where the pulling force
is applied. For the sake of conciseness we present in Fig. \ref{fig:experiment} the results for the case where the pulling force and torque were applied as schematically represented in Fig. \ref{fig:tripod}.
Negative values of $F$ correspond to the case when the force was applied  in the diametrically opposed direction.

\section{Theoretical Predictions}
\label{TheorySection}
%
\begin{figure}
\begin{center}
\includegraphics[width=85mm]{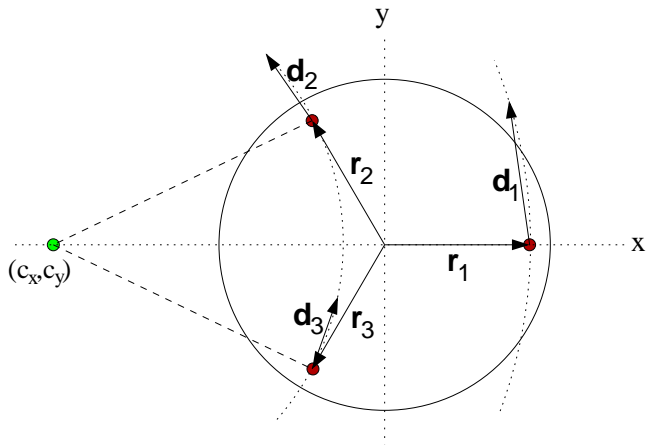}
\end{center}
\caption{Geometry for the calculation of the critical torque and force for a tripod
(see text).}
\label{fig:tripodcalculation}
\end{figure}

Let us consider a tripod of mass $M$ which rests on three symmetrically arranged legs, as sketched in Fig.~\ref{fig:tripodcalculation}. Using a coordinate system with origin at the center of the tripod, these legs are located at unit distance $R=1$ at
\begin{equation}
\rvec_i=(\cos\theta_i,\sin\theta_i,0)\,, \qquad i=1,2,3
\end{equation}
where
\begin{equation}
\theta_1=0, \quad \theta_2=2\pi/3,\quad \theta_3=4\pi/3\,.
\end{equation}
The mass distribution of the tripod is symmetric, i.e., in the absence of external forces
all legs carry the same weight.

\begin{figure}
\begin{center}
\includegraphics[width=60mm]{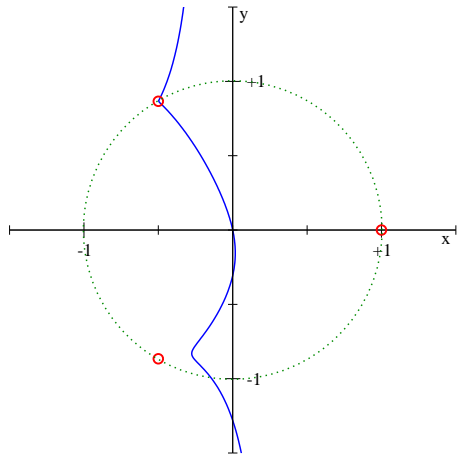}
\end{center}
\caption{Possible position of the coordinates $c_x,c_y$ (solid line) for  different values of the critical pair $F,T$ under the condition $F_y=0$, i.e., for a horizontal force. The dotted circle is the tripod's perimeter while the small circles represent the tripod's feet. The asymptotes $y \to \pm\infty$ correspond to the case of a pure translation ($T=0$), which can be interpreted as a rotation around a point at infinity. Note that for a particular pair $(F,T)$ the vector $\cvec$ coincides with the position of the upper foot of the tripod. This special situation accounts for the linear segment of the $F$-$T$ curve on the left--hand side of Fig. \ref{fig:experiment}. The point where $\cvec=0$ corresponds to the case
of a pure torque ($F=0$).}
\label{fig:rotation}
\end{figure}

As shown in Fig.\ref{fig:tripod}, the tripod is subjected to external forces, namely, a linear force $\Fvec_{\rm ext}$ applied at the center of the disk and at height $h$ above the track, as well as an external torque $\Tvec_{\rm ext}$ in form of a pair of two oppositely oriented lateral forces $\mathbf K$ and $-\mathbf K$. We assume the tripod to be perfectly rigid. Before sliding sets in, the forces will
displace and rotate the tripod infinitesimally, leading to individual microscopic displacements $\dvec_1,\dvec_2,\dvec_3$ of the contact points. Generally the displacement of the disk is a combination of a translation and a rotation with respect to the origin which -- according to a well-known theorem in classical mechanics -- may be viewed as a \textit{pure} rotation by an angle $\delta\omega$ around some point
\begin{equation}
\cvec = (c_x,c_y,0)\,.
\end{equation}
This point may lie inside or outside the perimeter of the tripod (see Fig. \ref{fig:rotation})
and can be used to parametrize the possible displacements at the contact points
\begin{equation}
\dvec_i=\delta\wvec \times (\rvec_i-\cvec)\, \qquad i=1\ldots 3
\end{equation}
where $\delta\wvec=(0,0,\delta\omega)$. Note that in the case of a sliding disk
studied in~\cite{Zeno2004}, the diplacements are controlled by only one
parameter $\epsilon$ (see Eq. (\ref{eq:epsilon})) because of rotational symmetry.
The same applies for the static case \cite{Silvio2004}, where $\epsilon$ has to be replaced by a parameter
$\gamma= \frac{|\cvec|}{R}$. The tripod, however, is no longer rotationally symmetric
and thus we have to introduce {\it two} parameters $(c_x,c_y)$.

As outlined in the Introduction, any theory of static friction has to make assumptions about the effective microscopic mechanism that compensates the external forces before sliding sets in. Here we use the \textit{collective breaking scenario} discussed previously ~\cite{Silvio2004}. In this scenario it is assumed that each bolt establishes microcontacts
with the track, whose number is proportional to the normal force.
Each of these microcontacts can be thought of as an elastic spring. The microscopic displacement $\dvec_i$ deforms these springs, generating a certain restoring force in the $xy$ plane. For small displacements these restoring forces are assumed to follow Hooke's law. However, when the restoring force at a spring
reaches a certain threshold, the microcontact breaks and immediately reattaches to the surface, redistributing its stress among the other microcontacts. This leads to a microscopic stick--slip creeping motion by which the stress is continuously redistributed among the microcontacts until all of them sustain on average the same stress
(even if their individual displacements
$\dvec_i$ are different). Sliding sets in as soon as this redistributed average stress exceeds the threshold \textit{simultaneously} at all contact points. This critical state at the onset to sliding defines a curve in the $F$-$T$ diagram. The shape of this curve is nontrivial and reveals information about the microscopic mechanism at the transition from static to sliding friction.

To determine this critical curve let us introduce the normal forces at the contact points
\begin{equation}
\Nvec_i = (0,0,-N_i).
\end{equation}
According to the \textit{collective breaking scenario} described above the displacements of the contact point at the onset of sliding induce a restoring force
\begin{equation}
\Rvec_i=-\mu_s N_i \evec_i\,,
\end{equation}
where $\evec_i=\dvec_i/|\dvec_i|$ are the unit vectors along the direction of displacement (in which $\delta\omega$ cancels out) and $\mu_s$ is the static
friction coefficient. The resulting individual forces at the three legs
\begin{equation}
\Fvec_i=\Nvec_i+\Rvec_i
\end{equation}
give rise to a total restoring force and a restoring torque
\begin{equation}
\Fvec_{\rm rest}=\sum_{i=1}^3 \Fvec_i \,, \qquad
\Tvec_{\rm rest}=\sum_{i=1}^3 \rvec_i \times \Fvec_i \,.
\end{equation}

At the threshold of static friction, i.e. immediately before sliding sets in, these forces have to balance the external forces and weight of the disk, leading to the equations
\begin{equation}
\label{eqa}
\Fvec_{\rm rest}+\Fvec_{\rm ext}-Mg\evec_z=0
\end{equation}
and
\begin{equation}
\label{eqb}
\Tvec_{\rm rest}+\Tvec_{\rm ext}+\hvec\times\Fvec_{\rm ext}=0\,,
\end{equation}
where $\Tvec_{\rm ext}=(0,0,\tau)$ is the torque applied at a height $h$ above the track and
$\Fvec_{\rm ext}=(F_x,F_y,0)$ is the external force applied at the point $\hvec=(0,0,h)$.
Eqs. (\ref{eqa}) and (\ref{eqb}) form a system of six equations. Choosing a rotation center $\cvec=(c_x,c_y,0)$ these equations allow one to determine the six unknowns $N_1,N_2,N_3,F_x,F_y$, and $\tau$.

Three of the six equations, namely, the third component of Eq. (\ref{eqa}) as well as the first and second components of Eq. (\ref{eqb}), are linear and can be solved in order to determine the normal forces. The result reads
\begin{eqnarray}
\label{normalforces1}
N_1 &=& \frac13 \left( Mg + 2hF_x \right) \\
N_2 &=& \frac13 \left( Mg + h(\sqrt{3}F_y-F_x ) \right) \\
\label{normalforces3}
N_3 &=& \frac13 \left( Mg - h(\sqrt{3}F_y+F_x ) \right)\,.
\end{eqnarray}
As can be seen, the external force $F_{\rm ext}$ applied at nonzero height $h>0$ leads to an additional torque and hence to different normal forces at the three bolts. As expected, for $h=0$ one obtains $N_1=N_2=N_3=Mg/3$.

Inserting the solution for the normal forces (\ref{normalforces1})-(\ref{normalforces3}) into Eqs.~(\ref{eqa}) and (\ref{eqb}) three nonlinear equations remain to be solved. Solving the first two components of (\ref{eqa}) with \textit{Mathematica} one obtains a highly complex non-linear expression for $\Fvec_{\rm ext}$ and similarly, by inserting this solution into the third component of Eq.~(\ref{eqb}), an expression for $\tau$ as a function of $\cvec=(c_x,c_y,0)$. Therefore, scanning all possible values of $c_x$ and $c_y$ for which $N_i\geq 0$, one obtains the critical curve $F=|\Fvec_{\rm ext}|$ versus $T=|\Tvec_{\rm ext}|$.

This curve is ploted in Fig. \ref{fig:experiment} as a solid line along with the experimental data, assuming that
the pulling force points in horizontal direction. One may notice in this figure that there is an asymmetry
in the data with respect to reflections $F\to -F$, which is confirmed by our numerical solution. The reason is that
the external force is applied at height $h>0$, imposing an additional torque leading to different normal forces.
Interestingly, this gives rise to a regime on the l.h.s. of Fig.~\ref{fig:experiment}, where the coordinates ($c_x,c_y$)
coincide with the location of one of the tripod's feet, meaning that the whole tripod microscopically rotates
about this pivoting point (see Fig. \ref{fig:rotation}). It turns out that this happens in a noticeable range
of values of $(F,T)$ for which the values of $c_x$ and $c_y$ happen to be exactly under the foot. This corresponds
to the region of linear dependence between force and torque. We note that for $h=0$ the theory would predict a symmetric curve without such a linear segment.


\section{Conclusions}
%
In this paper we studied both experimentally and theoretically the coupling
between static friction and torque for a tripod in dry
contact with a track. Our results indicate that there is a nontrivial
coupling between $F$ and $T$. These results are of relevance particularly
in the field of programmed motion, where one needs to program into a
machine the exact forces and torques to get some task accomplished (like moving a heavy object across the floor).

Recent experiments by Rubinstein and coworkers have shown the role
of propagating fronts of broken microcontacts at the transition from
static to sliding friction \cite{Rubinstein04}. However, due to the
inhomogeneity of their experimental setup it is not clear how important
a role elasticity plays.
In order to assess this question one could think of repeating the
photoarray experiment with a cylinder or a tripod
subjected simultaneously to a torque and a pulling force. In this case
each microcontacts are characterized
by different displacements so that the elasticity of the body should
play a minor role.

%
\section*{Acknowledgements}

S.R.D. would like to thank the warm hospitality at the Institut f\"ur Physik
at the University of W\"urzburg and at the Computational Physics Group at the University of Duisburg--Essen.
\appendix

\section{The special case $\theta_1=\pi/6, h=0$:}
%
\setcounter{equation}{0}

Analyzing the equations that determine the critical curve for a general angle $\theta_1>0$ between the first
bolt and the $x$-axis in the special case of zero height $h=0$ we realized that the choice
\begin{equation}
\theta_1=\pi/6, \quad \theta_2=5\pi/6,\quad \theta_3=9\pi/6
\end{equation}
is special in so far as the vector pointing to the center of rotation $\cvec$ and the
total restoring force $\Fvec_{\rm rest}$ are always orthogonal and aligned with the coordinate system, i.e.,
\begin{equation}
c_x=0\,, \quad F_y=0\,.
\end{equation}
In this case the equations~(\ref{eqa}) and (\ref{eqb}) can be simplfied and solved analytically.
The result -- now parametrized by a {\em single} parameter $c_y$ -- reads
\begin{eqnarray}
F_x &=& \frac13 \left( \frac{1-2c_y}{\sqrt{1-c_y+c_y^2}}-sgn(1+c_y)\right) \\
|\Tvec| &=& \frac13 \left( \frac{2-c_y}{\sqrt{1-c_y+c_y^2}}+sgn(1+c_y)\right) \,.
\end{eqnarray}
Depending on the sign of $1+c_y$ the parameter $c_y$ can be eliminated. In the resulting expression
one has to distinguish three different cases:
\begin{equation}
|\Tvec| = \left\{
\begin{array}{ll}
\frac12 \left( F_x+1+\sqrt{-2F_x-3F_x^2} \right) & \quad \mbox{if } -1 \leq F_x < \frac13(\sqrt{3}-1) \\[3mm]
\frac{2}{\sqrt{3}}-F_x & \quad \mbox{if } \frac13(\sqrt{3}-1) \leq F_x \leq \frac13(\sqrt{3}+1) \\[3mm]
\frac12 \left( F_x+1+\sqrt{-2F_x-3F_x^2} \right) & \quad \mbox{if } \frac13(\sqrt{3}+1) < F_x \leq 1
\end{array}
\right.
\end{equation}
In the first and third case all bolts of the tripod start sliding while in the second case the tripod starts to rotate around the bolt located at $\rvec_3=(0,-1,0)$.


\end{document}